\documentstyle[emulateapj,psfig]{article}

\newcommand {\BS}{{{\it Beppo}SAX}\ \ignorespaces}
\def\gta{ \lower .75ex \hbox{$\sim$} \llap{\raise .27ex \hbox{$>$}} }
\def\lta{ \lower .75ex\hbox{$\sim$} \llap{\raise .27ex \hbox{$<$}} }

\begin{document}

\title{Testing Comptonization models using\\
 {\it Beppo}SAX observations of Seyfert 1 galaxies}
\author{P.O. Petrucci\altaffilmark{1,2}, F. Haardt\altaffilmark{2},
 L. Maraschi\altaffilmark{1}, P. Grandi\altaffilmark{3},
 J. Malzac\altaffilmark{1}, G. Matt\altaffilmark{6}, 
 F. Nicastro\altaffilmark{3,4,5},
 L. Piro\altaffilmark{3}, G.C. Perola\altaffilmark{6}, A. De
 Rosa\altaffilmark{3}.}

\altaffiltext{1}{Osservatorio Astronomico di Brera, Milano, Italy} 
\altaffiltext{2}{Universit\'a dell'Insubria, Como, Italy}
\altaffiltext{3}{IAS/CNR, Roma, Italy} 
\altaffiltext{4}{CfA, Cambridge Ma., USA}
\altaffiltext{5}{Osservatorio Astronomico di Roma, Roma, Italy}
\altaffiltext{6}{Universit\'a degli Studi ``Roma 3'', Roma, Italy}

\begin{abstract}
We have used realistic Comptonization models to fit high quality BeppoSAX
data of 6 Seyfert galaxies. Our main effort was to adopt a Comptonization
model taking into account the anisotropy of the soft photon field. The
most important consequence is a reduction of the first scattering order,
which produces a break (the so-called anisotropy break) in the outgoing
spectra. Thus anisotropic Comptonization models yield spectra with convex
curvature. The physical parameters of the hot corona (i.e. the
temperature and optical depth) obtained fitting this class of models to
broad band X-ray spectra are substantially different from those derived
fitting the same data with the power law + cut--off model commonly used
in the literature. In particular, our best fits with Comptonization
models in slab geometry give a temperature generally much larger and an
optical depth much smaller than derived from the power law + cut--off
fits, using standard Comptonization formulae. The estimate of the
reflection normalization is also larger with the slab corona model.  For
most objects of our sample, both models give Compton parameter values
larger than expected in a slab corona geometry, suggesting a more
``photon starved'' X-ray source configuration.  Finally, the two models
provide different trends and correlation between the physical parameters:
for instance, with the power law + cut-off fits, we obtain a correlation
between the reflection normalization and the corona temperature whereas
we find an anti-correlation between these parameters with the slab
geometry. These differences have major consequences for the physical
interpretation of the data. In the framework of reprocessing models, the
cut-off power law best fit results suggest the thermal corona to be
dominated by electron-positron pairs. On the contrary, the slab corona
model is in better agreement with a low pair density solution.
\end{abstract}

\keywords{galaxies: active; galaxies: Seyfert; X-rays: galaxies}

\section{Introduction}
The broad band X-ray spectra (2-300 keV) of Seyfert 1 galaxies are
generally well fitted by a cut-off power law continuum with superimposed
secondary components like a neutral iron line and a reflection hump. The
latter components have been detected in the end of the eighties with
GINGA (Piro et al. \cite{pir90}; Matsuoka et al. \cite{mat90}; Pounds et
al. \cite{pou90}) and are considered as signatures of reprocessing of
the primary X-ray emission in surrounding cold matter. A high energy
cut-off in the X-ray/$\gamma$-ray part of the spectrum was first
unambiguously detected by SIGMA and OSSE in NGC 4151 (Jourdain et al.,
\cite{jou92}; Maisack et al. \cite{mai93}). A high energy cut-off was
also requested to fit the average X-ray/$\gamma$-ray spectra (obtained
from data of different satellites) of a sample of Seyfert (Zdziarski et
al. \cite{zdz95}; Gondek et al. \cite{gon96}). The broad band
capabilities of the Italian-Dutch satellite BeppoSAX allowed to detect
high energy cut-offs in other Seyfert 1 galaxies, about half of the
fifteen observed with this satellite. Values of the high energy cut-offs
range from 70 keV to $\sim$ 300 keV (Matt \cite{mat00}, hereafter
M00).\\

A cut-off power law plus reflection model (the so-called PEXRAV model in
XSPEC, Magdziarz \& Zdziarski \cite{mag95}) depends on three parameters:
the spectral index $\Gamma$, the high energy cut-off $E_c$ and the
reflection normalization $R$, the formers characterizing the shape of the
primary continuum. The physical interpretation of these parameters is
generally done in the framework of thermal Comptonization mechanism,
which is commonly believed to be the origin of the X-ray emission of
Seyfert galaxies. Using approximate relations (cf. section \ref{result}),
it is then possible to derive, from values of $\Gamma$ and $E_c$, values
of the temperature $kT_e$ and optical depth $\tau$ of the
``comptonizing'' hot plasma (the so-called corona).\\

If such approximations are sufficient in isotropic geometries (e.g. a
central soft photon source surrounded by a spherical hot cloud), where
Comptonization process produces roughly power law spectra with a high
energy cut-off, strong discrepancies may appear in anisotropic ones,
especially for small optical depth and large temperature. Indeed, the
(possible) anisotropy of the source of seed photons (as seen by the
corona), may introduce anisotropies and modifications of the outgoing
spectrum. The largest effect occurs when soft photons are emitted by a
plane (disk) below the corona. In this case, photons backscattered
towards the disk in the first scattering are necessarily produced in
"head--on collisions" and therefore have an energy gain larger than
average, while photons scattered towards the corona (i.e. in the forward
direction) have an energy gain smaller than average.  As a consequence,
the contribution of the first scattering order to the outgoing flux is
significantly reduced producing a spectral break (the so-called
anisotropy break) in the spectrum. Below the high energy cut--off the
spectra is then better described by broken power laws than by a simple
power law contrary to what is generally believed (Petrucci et al.,
\cite{pet00}, hereafter P00). These anisotropic effects become however
less important for larger optical depth (i.e. $\tau>1$) since they are
``smoothed'' by the larger number of Compton scatterings suffered by the
soft photons.\\

It is also worth noting that the energy of the anisotropy break which is,
in practice, close to the peak energy of the third scattering order
(Haardt \cite{haa93}), depends on $kT_e$ but also on the temperature of
the seed soft photon plasma $kT_{bb}$. Consequently, if in the case of
isotropic geometry the spectral shape is relatively independent on
$kT_{bb}$, which fixes only the low energy boundary of the emitted
spectrum, in anisotropic geometry the shape of the Comptonization spectra
depends on it.\\

In a recent paper (P00), we have applied accurate Comptonization
models to fit high quality data of the \BS long look at NGC~5548. We
have underlined the importance of the anisotropic effects and, in
particular, we have shown that due to the presence of the anisotropy
break, above which the intrinsic continuum is steeper, the temperature
of the hot electrons estimated from Comptonization models in
anisotropic geometry can be much larger that the one derived fitting
the data with a power-law + cut off model.\\

Besides, we have shown that these different models may provide
different (and even opposite) relationships between physical
parameters. For instance, still in the case of NGC~5548, the spectral
softening of the spectrum, observed during a small flare in the
central part of the observation, goes with an increase of the high
energy cut--off (i.e. of the corona temperature), in the case of
PEXRAV, but with a decrease of the corona temperature when using an
anisotropic Comptonization model.\\

In the same spirit of P00, we extend our application of accurate
Comptonization models to a larger sample of objects. We report in this
paper the results of this study. The paper is organized as follows. In
section \ref{data} we briefly present the main results of the BeppoSAX
observations of Seyfert galaxies. We also present the sample of objects
we use for the purpose of this paper. We describe the models we used and
the fitting procedure in section \ref{modfit}. We also detail in this
section the fit of the soft excess/warm absorber (hereafter WA) features
observed in each object. We present the results of the fits in section
\ref{result} and discuss their physical interpretations in section
\ref{discu}. We then conclude in the last section.

\section{The data}
\label{data}
\subsection{Overview of the \BS observations of Seyfert I galaxies}
Two major \BS Core programs were devoted to classical Seyfert 1
galaxies aiming to study the broad band spectra (PI: G.C. Perola) and
spectral variability (PI: L. Piro). Thirteen sources have been
observed on aggregate so far, since the launch of the satellite in
1996.\\

Reviews summarizing the main results on these sources have been presented
by Piro et al., (\cite{pir99}) and M00. For all objects, the spectral
models used by the authors included a power law with exponential cut--off
(i.e. $\displaystyle f_{E} \propto E^{1-\Gamma}\, {\rm e}^{-E/E_{\rm
c}}$) characterized by a photon index $\Gamma$ and a e--folding energy
$E_{\rm c}$, a gaussian to fit the neutral Iron line near 6.4 keV and a
Compton reflection continuum characterized by a reflection normalization
$R$. Soft excess and/or WA features were added if required by the data.\\

The major findings are the followings:
\begin{itemize}
\item A high energy cut--off is generally observed, with an
$e$-folding energy typically in the range 70--300 keV.
\item All sources show the presence of a reflection component, so
confirming the GINGA and ASCA results (Nandra \& Pounds,
\cite{nan94}). 
\item Only two objects (Mkn 509 and Mkn 841) show the presence of a soft
excess. This sharply contrasts with the results obtained with EXOSAT
(Turner \& Pounds \cite{tur89}) or ROSAT (Walter \& Fink, \cite{wal93})
where most of the Seyfert 1 observed seemed to show such excess (for a
discussion of this point see Piro et al., \cite{pir97}). On the other
hand, WA features (mostly the O VII and O VIII edges) are generally
observed in the LECS range.
\item A correlation between the power law index $\Gamma$ and the amount
of reflection component $R$ is indicated, in the same sense as found by
Zdziarski et al. (\cite{zdz99}, hereafter Z99). However, the issue of the
reality of this correlation is still open since it may be due to the fact
that the two parameters are strongly related in the fitting procedure
(see for example Vaughan \& Edelson \cite{vau00}).
\end{itemize}

\subsection{Our sample}
From the sample of Seyfert 1 galaxies observed by \BS, we have selected
observations with good signal-to-noise ratios at high energies as
measured with the PDS instrument. We therefore favored objects with hard
X-ray spectra.  Such observations are best suited for an analysis of the
high energy spectra.

The subsample of objects studied in this paper is reported in Table
\ref{tab1}, with the name of the different objects, the date of the
\BS observations (some of these objects have been observed several
times), the exposure time and the statistical significance in the PDS
instrument.\\
We also present results from the observation of ESO 141-G55, even if
its detection in the PDS instrument is relatively weak, since this
source was recently observed for the first time by \BS.\\

NGC 4151 was observed three times by \BS, two times in 1996 during the
SV-phase (in July and in December), and the last one in January 1999
during A02. However during the observation of July 96, the source
experienced a strong increase of the 2-10 keV flux by a factor two on
time scales of one day, with significant spectral variability as shown by
Piro et al. (\cite{pir00a}).  According to these authors, we choose to
study separately the low and high state occuring during this period (and
indicated as 96L and 96H in the following). Unfortunately, during the
high state, the LECS was off. Only the MECS and PDS data are therefore
available for this state. We have not used, in this paper, the
observation of December 96 since it does not differ substantially from
the low state of July 96.\\

Concerning Mkn 509, its observation has been split in two parts during
the AO2. Since no significant spectral variability was observed
between the two data sets (Perola et al. \cite{per00}), we will only
refer here to the combined spectrum.\\

In this paper, we will be concerned with data from three of the four
instruments on board \BS: the Low Energy Concentrator Spectrometer,
LECS (Parmar et al.  \cite{par97}) covering the 0.15--10 keV range,
the Medium Energy Concentrator Spectrometer, MECS (Boella et
al. \cite{boe97}) covering the 2--10 keV range and the Phoswich
detector system, PDS (Frontera et al. \cite{fro97}) covering the range
13 -- 200 keV. Due to some problems that remain in the spectral
analysis of LECS data above 4 keV, we shall use only data in the range
0.1--4 keV.\\

LECS and MECS event files and PDS pha files were downloaded from the
\BS SDC archives. The spectral counts were extracted from a circular
region of 4' and 8' radius around the source centroid in the MECS and
LECS images respectively. We used the data of the three (or two, for
observations done after May 1997) MECS units merged together to
increase the signal-to-noise ratio. In XSPEC (Arnaud \cite{arn96}),
we use the last updated responses (Sep. 1997) and background matrices
of each instrument.\\

\section{Model fitting}
\label{modfit}
\subsection{The primary continuum}
\label{primcont}
We fit the data using two different models for the primary continuum: 1)
an exponentially cut--off power law plus a reflection component from
neutral material (PEXRAV model of XSPEC, Magdziarzk \& Zdziardski
\cite{mag95}) and 2) a thermal Comptonization spectrum from a disk+corona
configuration in slab geometry. The latter was obtained using the code of
Haardt (\cite{haa94}, hereafter we will refer to this Anisotropic
Comptonization Code as AC2). This code derived the angle--dependent
spectra of the disk--corona system using an iterative scattering method,
where the scattering anisotropy was taken into account only in the first
scattering order. It includes also a reflection component described
following White, Lightman \& Zdziarski (\cite{whi88}) and Lightman \&
White (\cite{lig88}) and assuming neutral matter. The spectral shape of
the reflected photons is averaged over angles. It is multiplied by a
first normalization factor which depends on the inclination angle (see
Ghisellini, Haardt \& Matt \cite{ghi94} for details). In addition, the
usual $R$ normalization is left free to vary in the fit procedure, so
that, for the given inclination angle, $R=1$ corresponds to a solid
angle, subtended by the reflector, of 2$\pi$ . We do not apply
Comptonization models for other geometries (e.g. hemisphere) since, as
shown for NGC~5548 (P00), the results follow closely those obtained with
the slab geometry.\\

The fit parameters of AC2 are the temperature of the corona $kT_e$,
its optical depth $\tau$, the temperature of the disk $kT_{bb}$
(assuming a black body soft emission) and the reflection normalization
$R$. On the other hand, the PEXRAV continuum depends only on 3
parameters: the e--folding energy of the cut--off power law $E_{\rm
c}$, the photon index $\Gamma$ and the reflection normalization $R$.\\

\subsection{Soft X-ray excess and Warm Absorber}
\label{WA}
Most of the objects present WA features and/or a Soft excess in the
low part (i.e. in the LECS energy band) of their spectra. Good and
realistic fits require to take these features into account. Detailed
analyses of these different components are however beyond the scope of
this paper, which is instead focused on the high energy continuum of
the sources. Then, as a first approximation, we have added simple
components (edges, gaussian, blackbody...) to the primary continuum to
reproduce the main features present in the soft part of the
spectra. We discuss briefly the case of each source in the
following.\\

\subsubsection{NGC 4151}
\label{WA4151}
The \BS data of this galaxy have already been analyzed by Piro et al.
(\cite{pir00a}, \cite{pir00b}), assuming a simple cut--off power law
for the primary underlying continuum. A prominent soft excess, known
to exist in this object (Holt et al. \cite{hol80}) below the $\sim$2
keV cutoff due to a high absorbing column, is clearly detected in the
observations of 1996 (low state) and 1999. It is generally modeled by
allowing some fraction of the central source to be completely
uncovered or viewed through a lower column (the so-called ``leaky
absorber''). In addition, there is evidence of a possibly separate
soft excess component that does not vary with the 2--10 keV continuum
(Pounds et al., 1986; Perola et al., 1986). Part of this emission
comes from an extended X--ray region known to exist in this object
(Elvis et al. \cite{elv83}; Morse et al. \cite{mor95}). Recent
Chandra observations with the High Energy Transmission Grating
Spectrometer have revealed detail spectra of this source (Ogle et al.,
\cite{ogl00}) and allow for the first time precise measurements of the
ionization, temperature and kinetics of the extended soft X--ray
emission region.\\

For the scope of the present paper, we model the complex soft X--ray
emission in a (relatively) simpler manner. According to Piro et
al. (\cite{pir00a} \cite{pir00b}), we have used a dual absorber that is
the source is covered completely by a medium with a column density
$N_{H1}$ and with a covering fraction $f_{cov}$ by a second medium with a
column density $N_{Hcov}$. On the other hand, the soft excess below 2 keV
was modeled by a scattering component plus an ultrasoft component
described by a thermal bremsstrahlung, both absorbed by the galactic
column density.

Due to the lack of LECS data for the high state of 1996, we cannot
constrain the different components of the soft X-ray range. We thus
analyze the data only above 3.5 keV adding a simple photo-electric
absorption characterized by a column density $N_{H1}$.\\

We have reported for the three observations our best fit parameters
$N_{H1}$, $N_{Hcov}$ and $f_{cov}$, obtained with PEXRAV to model the
continuum, in Table \ref{pextable}.  We note a significant increase of
$N_{H1}$ between 96L and 99. As already noted by Piro et al
(\cite{pir00a}, \cite{pir00b}), such variations suggest that the
structure of the dual absorber has slightly changed between the two
observations.\\

\subsubsection{NGC 3783}
Evidences for deep absorption features in the 0.1-1.5 keV band of NGC
3783 have already been detected in ROSAT (Turner et al. \cite{tur93})
and ASCA observations (George et al. \cite{geo95}), the main ones
being the O VII and O VIII edges at 0.74 and 0.87 keV
respectively. More recently, in January 2000, NGC 3783 was observed
using the High Energy Transmission Grating Spectrometer on the Chandra
X-ray observatory (Kaspi et al.,\cite{kas00}). These authors detected
a large number of emission and absorption lines, strongly confirming
the presence of warm absorbing/emitting media embedding the central
X-ray source.\\

An emission feature near 0.5 keV has also been claimed in ASCA spectra
of this source (George et al. \cite{geo98}) and has been interpreted
as an O~VII emission line (expected at 568 eV).
Kaspi et al. (\cite{kas00}) clearly see this emission feature in the
Chandra spectra, confirming for the first time the detection of this
component.\\

According to these results, we have included in our models two
absorption edges, near 0.74 and 0.87 keV, and a gaussian peaking near
0.5 keV to fit the O~VII line. The best fit results are reported in
Table \ref{soft3783}. As suggested by the recent observation of
Chandra, the edge near 1.3--1.4 likely corresponds to contributions by
Fe L and Ne K edges whereas the second one, centered near 0.8--0.9
keV, may be due to the O VIII edge.\\

Concerning the line, we have checked that its addition significantly
improves the fit with $\Delta\chi^2$=16 (according to the F-test, this
difference implies a probability less than 0.01 that the line is due to
random fluctuations in the counts). This feature is thus detected by \BS
with a high confidence level. The corresponding best fit central energy
of the line was $E_{line}\simeq$0.55 keV with an EW of $\sim$100 eV.
Such EW is large, and it is unlikely that this emission features was only
produced by O VII emission lines. In fact, in the more extreme case where
we see the blends of the three O VII lines, we expect an EW of only 30 eV
(De Rosa, private communication). The large EW and small energy of the
line we obtained can be due in part to a bad modeling of the continuum
near this component.\\

\subsubsection{Mkn 509}
Mkn 509 was observed by BeppoSAX in 1998. The observation was split in
two parts during the AO2, the first one being done in May 1998 and the
second one in October 1998. As already said, a hardness ratio analysis
of the counts shows marginal evidence of spectral variations between
the two observations. They have thus been combined and we applied the
spectral analysis to the integrated counts.\\

These data have already been studied in detail by Perola et
al. (\cite{per00}, hereafter PER00). Assuming a cut--off power law
model for the intrinsic continuum of this source, the data show at low
energy the presence of additional, superimposed features due to O VII
and/or O VIII edges as well as a strong soft excess. PER00 described
the latter by a ``soft'' power law component
($A_sE^{-\Gamma_s}$). They then obtained a significant improvement of
the goodness of the fit ($\Delta\chi^2$=20), the best fit values for
$A_s$ and $\Gamma_s$ being 1.11$\times 10^{-2}$ cm$^{-2}$ sec$^{-1}$
kev$^{-1}$ and 2.5 respectively.\\

PER00 underlined the fact that the addition of this ``soft'' component
had also the interesting effect of providing a more astrophysically
sound global solution, with a reflection normalization $R\sim 0.6$ and
Iron line width $\sigma_{Fe}\sim$ 0.4 keV (compared to $R\simeq$2 and
$\sigma_{Fe}\sim$3 keV without this soft component), giving some
confidence on the correctness of the model.\\

Following these authors, we have added this ``soft'' power law to our
fit, fixing $A_s$ and $\Gamma_s$ to the best fit values given
above. We also add an edge with the best fit energy and optical depth
found by PER00, i.e. $E_{edge}=0.74$ keV, $\tau_{edge}=0.06$.

\subsubsection{ESO 141-G55}
The presence of a soft excess was already suggested for this source from
ROSAT observations (Turner et al \cite{tur93a}). However the \BS data do
not show any evidence of such component nor WA features. It could be
partly due to the low S/N of this observation. Consequently, our fits
were done without any additional components in the soft band other than
the absorption by the galactic column density (cf. Tables \ref{pextable}
and \ref{h94tab2}).\\

\subsubsection{NGC 5548}
The WA observed in this source was already discussed in P00 and Nicastro
et al. (\cite{nic00}) and the reader can refer to these papers for more
details. For the purpose of this work, we adopt the results of P00 who
have added two edges, at $E_1$= 0.74 keV and $E_2$=0.87 keV, to fit the
Oxygen absorption features observed in the LECS band. The corresponding
optical depths were fixed to the best fit values obtained with PEXRAV,
i.e. $\tau_1$=0.51 and $\tau_2$=0.12 respectively.

\subsubsection{IC 4329A}
The X-ray spectrum of IC 4329A is known to display strong neutral
absorption, below 3 keV, due to a gas column about ten times greater
than the galactic value $N_{h_{gal}}=4.5\times 10^{20}$ cm$^{-2}$
(Elvis et al. \cite{elv89}). This gas is supposed to be associated
with the disk of the host galaxy, which is oriented nearly edge-on
(Petre et al. \cite{pet84}).\\

At low energies (below 1-2 keV), ROSAT and ASCA observations have also
shown evidence of two rather strong edges, one consistent with O VII,
the other with O VIII (Madejski et al. \cite{mad95}; Cappi et al.,
\cite{cap96}; Reynold \cite{rey97}; George et al. \cite{geo98a}).\\

The \BS observation of IC 4329A has been already analyzed by Perola et
al. (\cite{per99}) using PEXRAV as primary continuum. To model the soft
range of the spectrum, they added a neutral column of gas in addition to
the galactic one and included two absorption edges. This model gives a
good fit to the data and we adopt the same description in our fits. We
fix the energy and optical depth of the two edges and the neutral column
density to the best fit values obtained by Perola et al. (\cite{per99}),
i.e. $E_{1}=$0.73~keV, $\tau_{1}$=0.52, $E_{2}=$1.03~keV, $\tau_{1}$=0.19
and $N_h$=0.38$\times 10^{22}$ cm$^{-2}$.

\subsection{Fitting procedure}
Since the sample is only composed of Seyfert 1 galaxies, we expect, in
the unification model framework, relatively small inclination angles. In
the case of NGC 4151, there are claims of a large inclination angle
($\sim$60$^{\circ}$, Evans et al. \cite{eva93}). However this value is
rather uncertain. To reduce the length of the fitting procedure, we thus
decided to fix, for all the objects of our sample, the inclination to
30$^\circ$ (Nandra et al. \cite{nan97}).\\

The geometrical normalization of the reflection component $R$ was let
free to vary. We recall that a value $R=1$ corresponds to a covering
factor of the cold matter to the X--ray source of $\Omega=2\pi$ whatever
the degree of anisotropy of the X--ray emission.

We added a gaussian to reproduce the iron line complex near 6.4 kev.
The central energy and width of the gaussian were let free to vary.
An Iron edge in the range 7--9 keV was also added if required by the
data, that is for NGC 4151, NGC 3783.\\

We allowed the relative MECS to PDS normalization to vary by 5\%
around the value of 0.86 to account for the estimated systematic
uncertainty (Fiore et al. \cite{fio99}). Analogously, we let the LECS
to MECS normalization free to vary over the range of acceptable values
0.7--1 (Fiore et al. \cite{fio99}).\\

Finally, in the case of PEXRAV, we fixed the parameters of the soft
excess and/or WA components (i.e. edges, line, soft power law) to the
values presented in section \ref{WA}. For fits with AC2, we also
fixed, in a first step, the column density to the best fit values
obtained with PEXRAV. A second series of fits were obtained with a
column density free to vary. We checked that there were no significant
differences between the two sets of results (we generally obtain
slightly larger column densities and changes of the soft temperature
$kT_{bb}$ when $N_h$ is let free to vary). We will thus only consider
the best fit results with free column density in the following.\\

For computing errors on the parameters of the primary continuum (that is
$kT_e$, $\tau$, $kT_{bb}$ and $R$ for AC2, and $E_c$, $\Gamma$ and $R$
for PEXRAV), we let the column density free to vary but we fixed the
LECS/MECS and MECS/PDS normalization, and the Iron line and/or edges to
their best fit values. Throughout the paper, errors on single parameter
are quoted at a confidence level of 90 \% (i.e. $\Delta\chi^2$=2.7)
unless otherwise specified.\\

\section{Results}
\label{result}

\subsection{ The PEXRAV model}
\label{result1}
The best fit parameter values obtained with the fitting procedure
described above using the PEXRAV model are reported in Table
\ref{pextable}.  We checked, as far as possible, our PEXRAV results with
those already published on these data.  For IC 4329A, Mkn 509, NGC 4151,
NGC 3783 and NGC 5548 our PEXRAV fits are in good agreement with those
obtained by Perola et al. (\cite{per99}), Perola et al. (\cite{per00}),
Piro et al. (\cite{pir00a}, \cite{pir00b}), De Rosa et al. (\cite{der00})
and Nicastro et al. (\cite{nic00}) respectively. In the case of NGC 5548,
we also checked that a better modelization of the WA (with the CLOUDY
code used by Nicastro et al. \cite{nic00}) does not affect the results on
the continuum. We note also that, in the case of NGC 4151, we found
smaller errors for the spectral index $\Gamma$ ($\Delta\Gamma\simeq\pm
0.01$) in comparison to Piro et al. (\cite{pir00a}, where they found
$\Delta\Gamma\simeq\pm 0.1$). This is simply due to the fact that the
latters have computed the errors using the MECS and PDS data (above 3.5
keV) whereas we have used the LECS, MECS and PDS ones together (thus
adding the soft excess parameters in the fitting procedure).\\

For the PEXRAV model the parameters of the primary continuum determined
by the fits are $\Gamma$ and $ E_{\rm c}$.  The actual physical
parameters of the Comptonizing region (also reported in Table
\ref{pextable}) have been obtained from the spectral parameters in the
following way.  The temperature $kT_e$ is simply estimated as $kT_e\equiv
E_{\rm c}/2$, keeping in mind that such approximation roughly holds for
$\tau\lta 1$. For $\tau \gg 1$, $kT_e\equiv E_{\rm c}/3$ would be more
appropriate. Thus the reported values can be considered as upper limits
to the temperature. Knowing the temperature, the optical depth can be
computed from the spectral index derived from the PEXRAV fit using the
following relation (Shapiro, Lightman \& Eardley \cite{sha76}; Sunyaev \&
Titarchuk \cite{sun80}; Lightman \& Zdziarski \cite{lig87}):
\begin{equation}
\Gamma-1\simeq\displaystyle\left[\frac{9}{4}+\frac{m_ec^2}{kT_e
\tau(1+\tau/3)}\right]^{1/2}-\frac{3}{2} 
\label{eqtau}
\end{equation}
This equation is valid for $\tau> 1$, and we have checked {\it a
posteriori} that such condition is roughly matched in all cases.\\

\subsection{ The slab model}
The fits with the AC2 yield directly and consistently the temperature
$kT_e$ and optical depth $\tau$, which was our first motivation for
starting this approach. The best fit parameters are reported in Table
\ref{h94tab2}.  It is worth noting that another parameter enters the
fitting procedure in an important way, that is the temperature of the
soft photons, $kT_{bb}$. Its values (see Table \ref{h94tab2}) range, in
most cases, between 10 and 40 eV and are relatively well constrained with
errors of the order of 5-50\% whereas we have only data above 100 eV.
For large values of $kT_{bb}$ (i.e. $kT_{bb}>$30 eV), the high energy
tail of the black body shape may be detected in the {\it low energy part}
of the BeppoSAX data and $kT_{bb}$ can then be roughly constrained. We
recall that the intensity of the seed photon component with respect to
the comptonized one is also fixed by the fitted parameters (mainly
$\tau$).  When no tail is observed, the soft-X-ray data impose, at least,
an upper limit of $kT_{bb}$. For anisotropic geometries, the {\it high
energy part} of the Comptonization spectrum also depends on $kT_{bb}$,
through the location of the anisotropy break.  The fitting procedure has
thus to adjust $kT_{bb}$, together with $\tau$, $kT_e$ and $R$, to
reproduce the soft {\it and} hard X-ray data resulting in a complex
interdependence in the fitting procedure, between the soft and hard part
of the spectra.\\

It is worth noting that, in the case of NGC 4151 96L and 96H, the best
fit values of $kT_{bb}$ are relatively large (in fact we have only lower
limit, the AC2 working only for $kT_{bb}< 100$ eV). It is true however
that in the case of the high state 96H, we don't have any LECS
data. Consequently, as explained above, $kT_{bb}$ is not constrained by
the soft part of the spectra and large values may be reached. But it is
surprising that in the low state 96L, we still find large $kT_{bb}$, at
least 3 times larger than in 1999. A possible explanation is that, in the
observation of 1996, the presence of a larger reflection component (also
found with the PEXRAV fits) may hide the anisotropy break, thus
diminishing the constraints on $kT_{bb}$, and allowing larger values.

\subsection {Comparison between the slab model and PEXRAV}
A purely statistical comparison between the two models is not possible
since both give acceptable values of $\chi^2$. The slab model has one
more parameter and gives generally lower values of $\chi^2$, except for
NGC~4151. However, we have checked, for this source, that the best fit
$\chi^2$ probability are similar for the two models. Besides, the two
best fits have residuals without significant features and differences
between each other. The two models give thus an equivalent representation
of the data.

\subsubsection{The temperature and optical depth of the corona}
For all objects, the estimated corona temperatures (respectively
optical depths) from PEXRAV are substantially smaller (respectively
larger) than those inferred with an anisotropic Comptonization model.
This trend is similar to the the result obtained for NGC 5548 by
P00. In Fig. \ref{slabvspex}a we have plotted the corona temperatures
obtained with AC2 versus the temperatures obtained with PEXRAV. Large
differences (up to a factor 8) are found between the two
estimates. Note however that for IC 4329A, for which the high energy
data are comparatively better than for the rest of the sample, the two
models give results which are quite close to each other. \\

The reason of these differences is relatively simple. Within PEXRAV type
models, the slope of the power law, determined with small errors by the
LECS and MECS data, cannot change, by hypothesis, at higher energies. A
cut--off around 100 keV is then required to fit the PDS data. In
Comptonization models, the LECS and MECS data determine the slope below
the anisotropy break. Above this break the intrinsic spectrum is steeper
and thus can fit the PDS data without an additional steepening beyond 100
keV, allowing for a larger temperature and (consequently) a smaller
optical depth to keep, roughly, the same power law slope. \\


\subsubsection{The reflection component}
\label{reflec}
The reflection normalizations obtained with AC2 cluster around 1, except
for two extreme states of NGC 4151.  They are, in all cases but one (ES0
141-G55), larger than those found using the simple cut--off power-law
model. We have reported in Fig. \ref{slabvspex}b the amplitude of the
reflection component derived from both models. In some cases we obtain
differences of factors 4--5.  Note that with the exception of NGC~4151
for all other sources the slab model yields values of $R$ consistent with
1.\\

The case of NGC 4151 is peculiar, showing a large variation in $R$
between the 96 and 99 observations and an indication of a variation of
$R$ also between the low and high states of the July 96 observation.
These changes are present irrespective of the model used in the spectral
fitting (see also Piro et al. \cite{pir00a}, \cite{pir00b}).  However the
slab model yields higher absolute values, allowing for some reflection
also in the 99 observation for which PEXRAV yields $R=0.01\pm 0.1$.
According to the slab model, the corona temperature is very high in the
99 observation, which may offer a physical connection with the low value
of $R$ as dicussed in section \ref{corrsess}.\\ We can also remark that
the relatively small reflection component and the hardness of the
spectrum of this object may be difficult to reconcile with simple corona
geometries (like a plane or a patchy corona for example). More complex
ones (as dynamics coronae for example, Malzac et al. \cite{mal00}) may be
needed. Another possibility may be that the reflecting material in NGC
4151 is highly ionized. In this case, fitting with models which do not
take into account the complex ionization pattern of the reflector can
severely underestimate the reflection normalization (Ballantyne, Ross \&
Fabian \cite{bal00}; Done \& Nayakshin \cite{don01}; .\\

As an illustration of these results, we have plotted in
Fig. \ref{specpexh94}a and \ref{specpexh94}b the unfolded best fit
spectra of IC4329A and NGC 4151 (observation of 1999) obtained with
PEXRAV (dashed lines) and AC2 (solid lines). We have also plotted in each
case the reflection hump component to better estimate its contribution to
the total spectrum. In the case of IC 4329A, the two models give roughly
the same result at high energy. For NGC 4151 however, large differences
are expected between PEXRAV and AC2, with a factor of 2--3 at 300 keV and
larger than 10 above 500 keV.

\section{Discussion}
\label{discu}
A Comptonizing corona above a passive optically thick layer
(approximating an accretion disk with negligible internal heating) is a
complex system, in which the soft seed photons for the Comptonization
process are produced by reprocessing of the Comptonized emission in the
underlying passive layer. Such a corona is "self-cooled" and, for a given
geometrical configuration of the system, can only be in equilibrium if the
temperature and optical depth satisfy a definite relation (Haardt \&
Maraschi \cite{haa91}).  These relations which essentially correspond to
roughly constant Compton parameters $\displaystyle y \simeq
4\left(\frac{kT_e}{m_ec^2}\right)\,
\left[1+4\left(\frac{kT_e}{m_ec^2}\right)\right] \tau (1+\tau)$, with $y
\simeq$ 0.5 and 2 for the slab and hemisphere respectively, have been
accurately computed by Stern et al.  (\cite{ste95}; see also for a review
Svensson \cite{sve96}).

In Fig. \ref{figktetau}a and \ref{figktetau}b the values of $\tau$
vs. $kT_e$, obtained for the 6 sources considered here using the spectral
models AC2 and PEXRAV, are compared with the theoretical relations
expected for a Comptonizing region in energy balance for a plane (filled
rectangles, solid line) and hemispherical (filled hemispheres, dashed
line) configuration respectively.

The best fit results obtained with AC2 (Fig. \ref{figktetau}a) are
relatively close to the theoretical expectations for the slab case even
if they tend to fall preferentially above the solid line. The data
therefore indicate a Comptonization parameter larger than for a pure slab
geometry, that is a more ``photon-starved'' configuration. A similar
result was already derived by P00 for NGC 5548.  For the latter source,
best fits with a hemispherical Comptonisation model were also perfomed.
The derived coronal parameters were in fact in good agreement with the
theoretical predictions for the hemisphere. However an implausibly large
value of the reflection component (of the order of 2) was required,
suggesting that the real configuration is more complex than these two
ideal cases. For the present sample we did not perform fits with
hemispherical models.\\

The values of $kT_e$ and $\tau$ obtained with PEXRAV
(cf. Fig. \ref{figktetau}b) show the same trend as found with AC2 that is
larger values of $\tau$ for smaller values of $kT_e$. They also fall
above the theoretical expectation for a slab and even above the
theoretical expectation for a hemisphere (the theoretical curves of Stern
et al. \cite{ste95} have been extended to large $\tau$ assuming a
constant value of the Compton parameter for each curve i.e. y $\simeq$0.5
and 2 for the slab and hemispherical configuration
respectively). Therefore this set of parameters suggests, for each source
of our sample, a configuration more ``photon-starved'' than a
hemisphere.\\

An important difference with respect to the slab model is that the
optical depths derived from PEXRAV fits are generally larger than 1. The
corona should then be optically thick reducing or cancelling the effects
of anisotropy. Therefore also the PEXRAV model has an internal
consistency. However a corona with large optical depth may wash out
discrete features from the underlying disk (e.g. lines and reflection
itself, Petrucci et al. \cite{pet01b}) more than would be desirable. This
problem could be alleviated if the corona was "patchy" (and possibly
dynamic cf. section \ref{reflec}). \\

We conclude that both a hot, optically thin corona with significant
anisotropic effects and a less hot, optically thick, patchy corona
with negligible anisotropy are consistent with the available data.

\subsection{Correlations between physical parameters}
\label{corrsess}
A correlation between the reflection normalization $R$ and the photon
index $\Gamma$ has been claimed by Z99 from the study, using PEXRAV to
model the primary continuum, of a large number of GINGA observations of
Seyfert and galactic black hole objects. Yet, the validity of this
correlation is under debate since $R$ and $\Gamma$ are strongly
correlated in the fitting procedure (Vaughan \& Edelson \cite{vau00}). It
seems however more significant and less dispute in the case of the
galactic black hole objects where the errors on both parameters are
smaller.\\

We have plotted in Fig. \ref{Rvsgamma} the reflection normalization $R$
versus the photon index $\Gamma$ we obtained with PEXRAV.  In our case,
we do not find a clear correlation between the two parameters. A Spearman
rank-order correlation test gives inconclusive results with
$r_s\simeq$0.3. However our sample is biased in favor of objects with
hard spectra, to ensure a good detection in the PDS instrument. When the
13 objects actually observed by BeppoSAX are taken into account a
stronger correlation is observed (M00), in agreement with Z99.\\

In the case of the AC2 model we have no simple way of characterizing the
spectral shape. Also results are available for only 6 sources and we
cannot include the other Seyfert 1s observed by BeppoSAX.  Therefore a
direct comparison with the Z99 correlation is not possible. Below,
however, we examine the correlation of $R$ with the corona temperature
which can be computed for the two models.

\subsubsection{ Relation between $R$ or $\Gamma$ and the temperature of
the corona} 
\label{pexconc}
Z99 interpret the correlation between $R$ and $\Gamma$ in the framework
of thermal reprocessing models where $R$ is directly proportional to the
solid angle subtended by the cold matter surrounding the corona. The
latter solid angle is not necessarily proportional to the soft photon
flux reentering the corona, but it is likely that the larger is $R$, the
larger is the ratio of soft luminosity to hard luminosity in the corona,
i.e. the smaller is the Comptonization parameter $y$, resulting in softer
spectra (larger $\Gamma$).\\

It is interesting that the BeppoSAX data can provide the temperatures
associated with sources with different values of $R$ or $\Gamma$, which
were not available from the GINGA data used by Z99. The result is shown
in Fig. \ref{RvsEc} where we plot $R$ vs.  the corona temperature $E_c/2$
obtained with PEXRAV.
Here again, in order to have a larger sample, we added the best fit
values obtained by M00 for the other Seyfert 1s observed with BeppoSAX.
This plot suggests a {\it positive} correlation between the two
parameters, that is, the temperature of the corona is {\it larger} for
{\it larger} values of the reflection normalization.  In fact a plot of
$E_C/2$ vs. $\Gamma$ shows that steeper spectra correspond to higher
temperatures and (as a corollary) to lower optical depths
(cf. Fig. \ref{gammavsTe}a and \ref{gammavsTe}b).\\

The slab model analysis yields different trends. A plot of $R$ vs. $kT_e$
for the latter model shows that sources with {\it larger} $R$ tend to
have {\it lower} $kT_e$ (cf. Fig. \ref{RvskTe}), a behaviour opposite to
what is found with PEXRAV. The suggested correlation depends however
strongly on the different states of NGC~4151.\\

To summarize, an analysis in terms of PEXRAV indicates the following
correlations: {\it larger} reflection component (steeper $\Gamma$) --
{\it larger} temperature (cf. Fig. \ref{RvsEc}) while the analysis in
terms of anisotropic Comptonization yields: {\it
  larger} reflection -- {\it lower} temperature (cf. Fig. \ref{RvskTe}).\\

The latter behaviour is naturally expected in pair free Compton cooled
coronae (see Fig \ref{figktetau}a) since the transition to lower Compton
parameter (i.e. higher $l_s$/$l_h$ as suggested by a larger $R$) implies
a decrease in the temperature for constant optical depth, as (relatively
well) verified by our sources.\\

The behaviour indicated by PEXRAV could instead be understood better in
the case of a pair dominated corona in pair equilibrium.  In such a case,
for constant hard compactness, the larger the ratio $l_s/l_h$, the
larger the temperature of the corona.  This behavior is simply imposed by
the pair equilibrium. Indeed, an increase of the cooling corresponds to a
decrease of the number of particles in the hard tail of the thermal
particle distribution. To reach a new equilibrium the temperature must
increase.  (Zdziarski \cite{zdz85}; Svensson \cite{sve82}; Ghisellini \&
Haardt \cite{ghi94b} ; Coppi \cite{cop99}).\\


A possible limitation of the ``pair dominated'' interpretation comes from
the high values of the compactness $l_h$ required by the physical
parameters determined with PEXRAV. We have reported on
Fig. \ref{figktetau2} the $\tau$--$T_e$ relations expected, for a pair
dominated corona, for different values of $l_h$, but varying the ratio
$l_h/l_s$. We have also included the observations of M00. The data
require values of $l_h$ in the range 100--1000. Given the (unabsorbed)
luminosity of the sources of our sample ($L_{0.1-200\ keV}$ of the order
of $10^{44}$ erg.cm$^{-2}$.s$^{-1}$), this implies black hole mass upper
limits of the order of $10^7$ solar masses (for $l_h$=100).  This is
relatively smaller than (but roughly consistent with) black hole mass
estimations obtained from reverberation mapping or resolved kinematics
methods for most of the objects of our sample (see Gebhardt et
al. \cite{geb00} and Nelson \cite{nel00} and references therein). The
worse case is that of NGC~4151, where the PEXRAV best fit values of
$kT_e$ and $\tau$ require very large hard compactness (larger than
1000), implying a black hole mass of $\sim 10^5$ M$_{\sun}$
(reverberation mapping methods estimate a black hole mass of $\sim 10^{7}
M_{\sun}$, Clavel et al. \cite{cla87}; Ulrich \& Horne \cite{ulr96}). In
this case, however, the complexity of the soft part of the spectrum may
lead to an uncorrect modelisation of the continuum. We therefore conclude
that the pair dominated coronae interpretation is also still valid for
this object. Other observations (with better constraints on the
compactness and the physical parameter of the corona) are clearly needed
to confirm or disprove these results.

\section{Conclusion}
The aim of this paper was to test anisotropic Comptonization models over
the high signal to noise \BS observations of a sample of six Seyfert 1
(seven observations). We use two types of model: a detailed
Comptonization code in slab geometry, which treats carefully the
anisotropy effects in Compton processes, and a simple cut--off power law
plus reflection model (PEXRAV model of XSPEC). The latter is generally
used as a zero order approximation to Comptonization spectra. If this is
a relatively good approximation for isotropic geometry, it fails to
reproduce the real shape of Comptonization spectra in anisotropic ones
(like a plane, hemispherical or spherical hot region above a flat
disk). In this case, due to the deficiency of the first order scattering
component towards the observer, it better resembles a broken power law
with convex curvature. This leads to strong differences between the best
fit parameter values obtained with these two models. The main results of
this work are the following:
\begin{itemize}
\item The data are well fitted by both models and there is no
statistical evidence for a model to be better than the other. Both
models give results in agreement with a X-ray source geometry more
``photon starved'' than the slab case.
\item Our best fits with Comptonization models in slab geometry give a
temperature generally much larger and an optical depth much smaller than
derived from the power law + cut--off fits, using standard Comptonization
formulae. The estimate of the reflection normalization is also larger
with the slab corona model.
\item The two models also lead to different relationships between
physical parameters. For instance, PEXRAV indicates a correlation between
the reflection normalization and the corona temperature whereas the slab
corona model suggests an anticorrelation. This has major consequences for
the physical interpretation of the data. Thus, the PEXRAV results suggest
that the hot corona may be pair dominated whereas the slab corona model
is in better agreement with a low pair density solution.
\end{itemize}

These results should have observational consequences. First, the large
differences in the fitted temperatures lead to widely different
predictions as to the fluxes emitted at higher energies, above the PDS
range. This underlines the need of better soft $\gamma$--ray
observations, as those possibly provided by INTEGRAL, in order to
directly confirm or disprove the temperatures inferred with anisotropic
Comptonization models.\\

On the other hand, the large values of $\tau$ predicted by PEXRAV
would spread out the reflection hump in the entire X-ray range,
avoiding any detection of this component. The fact that we obtain not
negligible values of $R$ with PEXRAV may however be reconciled with
the large values of $\tau$ if the corona is ``patchy''. Indeed, in
this case, reflection could come from the uncovered part of the cold,
reflecting material. Part of the reflection could also come from the
torus. The variability of the reflection component may allow to
differentiate these two possibilities. Large values of $\tau$ would
also modify the profile of the Fe line component produced in the disk,
leading to an attenuation and broadening of the line. Forthcoming
observations with CHANDRA and XMM-Newton are expected to bring
substantial progress on this issue.\\

\noindent
{\sl Acknowledgements:} POP acknowledges a grant of the European
Commission under contract number ERBFMRX-CT98-0195 (TMR network
"Accretion onto black holes, compact stars and protostars"). This work
was partially supported by the Italian MURST trough the grant
COFIN98-02-15-41 (JM) and by the Agenzia Spaziale Italiana (ASI) trough
the grant ASI-ARS-99-74 (LM,FH).

\clearpage
\begin{table}
\begin{tabular}{llccc}
\tableline 
\tableline 
Source& Observation& Exposure&PDS& References\\
Name& Date& Time (ks)& Detection&\\
\tableline
NGC 5548& Aug. 97 &300&55&Nicastro et al. \cite{nic00}\\
IC4329A& Jan. 98&100&68&Perola et al. \cite{per99}\\
NGC 4151& Jan. 99&100&182&Piro et al. \cite{pir00a}\\
& Jul. 96L&56&130&Piro et al. \cite{pir00a}, \cite{pir00b}\\
& Jul. 96H&15&90&Piro et al. \cite{pir00a}, \cite{pir00b}\\
ESO 141-G55&Nov. 99 &100&15&-\\
Mkn 509& May/Oct 98&100&29&Perola et al. \cite{per00}\\
NGC 3783&Jun. 98 &300&55&De Rosa et al. \cite{der00}\\
\tableline
\end{tabular}
\caption{Sources names, observations dates and exposures of our
sample. We have also reported the statistical significance in $\sigma$ of
PDS detection in 13--300 keV band.  We have also reported, as far as
possible, the references presenting the first data analysis of each
observation.\label{tab1}}
\end{table}

\begin{table}[h]
\begin{tabular}{ccccccc}
\tableline 
\tableline 
$N_{H}$($10^{20}$cm$^{-2}$)\tablenotemark{a}&${E_{line}}^b$&$\sigma_{line}$&
${E_1}^b$&$\tau_1$&${E_2}$\tablenotemark{b}&$\tau_2$\\ 
\tableline
0.8$_{-0.6}^{+0.5}$&0.55$_{-0.1}^{+0.03}$&0.1$_{-0.04}^{+0.05}$&
0.84$_{-0.02}^{+0.02}$&0.63$_{-0.04}^{+0.09}$&1.31$_{-0.03}^{+0.10}$&
0.16$_{-0.04}^{+0.04}$\\
\tableline
\end{tabular}\\
\tablenotetext{a}{in excess of $N_{H, Gal.}=8.5\times 10^{20}$ cm$^{-2}$}
\tablenotetext{b}{in keV}
\caption{Best fit parameter of the soft X-ray complex of NGC 3783,
i.e. two edges and an emission line (see text for details), obtained
with PEXRAV. We have also reported the best fit column density, in
excess of the galactic one.\label{soft3783}}
\end{table}


\begin{table}
\begin{tabular}{lcccccc}
\tableline
\tableline
Source&$\Gamma$& $kT_e=\frac{E_c}{2}$ &$\tau$& R & $N_{h}$&$\chi^2$\\
\tableline 
NGC 5548&1.59$_{-0.02}^{+0.01}$&65$_{-15}^{+20}$&2.2$_{-0.4}^{+0.5}$&0.5$_{-0.1}^{+0.2}$&0.015$_{-0.001}^{+0.002}$&147/171\\
IC 4329 A&1.81$_{-0.02}^{+0.01}$&105$_{-20}^{+40}$&1.1$_{-0.3}^{+0.2}$&0.3$_{-0.1}^{+0.1}$&0.40$_{-0.02}^{+0.02}$&161/171\\
NGC 4151 99&1.45$_{-0.05}^{+0.15}$&45$_{-5}^{+25}$&3.4$_{-1.4}^{+0.6}$&0.01$_{0}^{+0.1}$&$N_{H1}=7.0_{-1.5}^{+0.5}\ N_{Hcov}=18_{-3}^{+2}$&177/179\\
&&&&&$f_{cov}=0.60_{-0.05}^{+0.05}$&\\
NGC 4151 96L&1.24$_{-0.01}^{+0.01}$&30$_{-3}^{+3}$&7.0$_{-0.8}^{+0.3}$&0.45$_{-0.05}^{+0.12}$&$N_{H1}=3.5_{-0.2}^{+0.9}\ N_{Hcov}=17_{-1}^{+3}$&106/83\\
&&&&&$f_{cov}=0.65_{-0.05}^{+0.05}$&\\
NGC 4151 96H&1.28$_{-0.11}^{+0.01}$&30$_{-2}^{+5}$&6.1$_{-0.7}^{+2.9}$&0.2$_{-0.1}^{+0.1}$&$N_{H1}=5_{-1}^{+1}$&61/53\\
ESO 141-G55&1.9$_{-0.05}^{+0.04}$&65$_{-25}^{+180}$&1.4$_{-0.9}^{+0.8}$&1.7$_{-0.6}^{+0.8}$&0.058$_{-0.005}^{+0.006}$&176/162\\ 
Mkn 509&1.60$_{-0.03}^{+0.03}$&40$_{-10}^{+15}$&3.1$_{-0.8}^{+0.8}$&0.7$_{-0.3}^{+0.3}$&0.048$_{-0.002}^{+0.001}$&163/202\\ 
NGC 3783&1.62$_{-0.02}^{+0.01}$&55$_{-10}^{+15}$&2.3$_{-0.4}^{+0.4}$&0.3$_{-0.1}^{+0.2}$&0.093$_{-0.006}^{+0.005}$&151/162\\ 
\tableline
\end{tabular}
\caption{Best fit parameters using the PEXRAV model: the photon index
$\Gamma$, the temperature $kT_e$ (in keV) and optical depth $\tau$ of
the corona, the reflection normalization $R$ and the best fit total
(i.e. intrinsic+galactic, in $10^{22}\ cm^{-2}$) column density for each
object. In the case of NGC 4151, the complex neutral absorber is
detailed in section \ref{WA4151}. The computation of $kT_e$ and $\tau$
are explained in section \ref{result1}.}
\label{pextable}
\end{table}


\begin{table}
\begin{tabular}{lcccccc}
\tableline
\tableline
Source&$kT_e$ & $\tau$ & k$T_{bb}$ & R & $N_{h}$ & $\chi^2$\\ 
\tableline
NGC 5548&230$_{-5}^{+10}$&0.18$_{-0.01}^{+0.01}$&22$_{-2}^{+3}$&0.9$_{-0.1}^{+0.1}$&0.027$_{-0.009}^{+0.011}$&144/170\\
IC4329A&170$_{-5}^{+10}$&0.20$_{-0.01}^{+0.01}$&15$_{-2}^{+3}$&1.0$_{-0.1}^{+0.3}$&0.35$_{-0.02}^{+0.01}$&158/170\\
NGC 4151 99&315$_{-5}^{+10}$&0.08$_{-0.01}^{+0.15}$&20$_{-1}^{+8}$&0.25$_{-0.05}^{+0.10}$&$N_{H1}=7.5_{-0.5}^{+0.5}\ N_{Hcov}=19_{-1}^{+1}$&184/178\\
&&&&&$f_{cov}=0.6_{-0.1}^{+0.2}$&\\
NGC 4151 96L&170$_{-10}^{+80}$&0.20$_{-0.10}^{+0.05}$&$>$60&1.8$_{-0.4}^{+0.9}$&$N_{H1}=4.3_{-0.6}^{+0.3}\ N_{Hcov}=20_{-5}^{+2}$&115/82\\
&&&&&$f_{cov}=0.65_{-0.05}^{+0.05}$&\\
NGC 4151 96H&190$_{-10}^{+40}$&0.18$_{-0.01}^{+0.11}$&$>$90&1.1$_{-0.1}^{+0.5}$&$N_{H1}=7.0_{-1}^{+0.5}$&66/52\\
ESO 141-G55&260$_{-30}^{+30}$&0.05$_{-0.01}^{+0.01}$&25$_{-3}^{+3}$&1.1$_{-0.3}^{+0.1}$&0.09$_{-0.02}^{+0.02}$&166/161\\
Mkn 509&210$_{-30}^{+20}$&0.17$_{-0.03}^{+0.06}$&38$_{-8}^{+7}$&1.0$_{-0.4}^{+0.6}$&0.07$_{-0.01}^{+0.01}$&164/201\\
NGC 3783&265$_{-5}^{+15}$&0.12$_{-0.01}^{+0.03}$&13$_{-3}^{+2}$&0.7$_{-0.3}^{+0.2}$&0.11$_{-0.02}^{+0.07}$&143/161\\
\tableline
\end{tabular}
\caption{Best fit parameters using the Comptonization model of H94: the
temperature and optical depth of the corona $kT_e$ (in keV) and $\tau$,
the soft photon temperature $kT_{bb}$ (in eV), the reflexion
normalization $R$ and the best fit total
(i.e. intrinsic+galactic, in $10^{22}\ cm^{-2}$) column density for each
object. In the case of NGC 4151, the complex neutral absorber is detailed
in section \ref{WA4151}.}
\label{h94tab2}
\end{table}

\clearpage
\vskip 1.5 truecm

\centerline{ \bf Figure Captions}

\vskip 1 truecm

\figcaption[]{(a) Best fit values of the corona temperatures obtained
with the cut--off power law + reflection model PEXRAV versus the best fit
values obtained with the Comptonization model in slab geometry H94. (b)
Same as (a) but with the best fit values of the reflection
normalization. The filled squares represent the different observations of
NGC~4151.\label{slabvspex}}

\figcaption[]{(a) Unfolded best fit spectra of IC 4329A with PEXRAV
(dashed lines) and H94 (solid lines). For each model, we have also
plotted the continuum without the reflection contribution and the
reflection hump. (b) Same as (a) but for the observation of 1999 of NGC
4151. \label{specpexh94}}

\figcaption[]{Optical depth $\tau$ versus temperature $kT_e$ from (a)
H94 and (b) PEXRAV fits. The theoretical relations between $\tau$ and
$kT_e$ for a plane and hemispherical Comptonizing region in energy
balance, for $kT_{bb}$=5 eV, are shown for comparison in solid and dashed
line respectively (filled rectangles and hemispheres are from Stern et
al. \cite{ste95}. These theoretical relations have been extended to large
optical depth assuming the constancy of the Compton parameter,
cf. section \ref{discu}). The filled squares represent the different
observations of NGC~4151.\label{figktetau}}

\figcaption[]{Reflection $R$ versus photon index $\Gamma$ from
PEXRAV. The open symbols are from M00 while the solid ones are the
results of this work. The filled squares represent the different
observations of NGC~4151.\label{Rvsgamma}}

\figcaption[]{Reflection component $R$ versus corona temperature $E_c/2$
from PEXRAV. The open symbols are from M00 while the solid ones are the
results of this work. The filled squares represent the different
observations of NGC~4151.\label{RvsEc}}

\figcaption[]{(a) PEXRAV corona temperature $E_c/2$ and (b) PEXRAV corona
optical depth $\tau$ versus photon index $\Gamma$. The open symbols are
from M00 while the solid ones are the results of this work. The filled
squares represent the different observations of
NGC~4151.\label{gammavsTe}}

\figcaption[]{Reflection normalization $R$ versus corona temperature
$kT_e$ obtained with the slab model. The filled squares represent the
different observations of NGC~4151.\label{RvskTe}}

\figcaption[]{Same as Fig. \ref{figktetau} but with extra $\tau$-$kTe$
curves expected in the case of a pair dominated spherical corona for
constant values of the hard compactness (in dot-dashed lines. From left
to right, $l_h$=1000, 100 and 10) but varying the starvness ratio
$l_s/l_h$. We have added the data from M00 in open symbols while the
solid ones are the results of this work. The filled squares represent the
different observations of NGC~4151.\label{figktetau2}}

\clearpage
\vskip 1.5 truecm
\centerline{ \bf Figures}
\vskip 1 truecm

\pagestyle{empty}
\setcounter{figure}{0}
\begin{figure}[h]
\plottwo{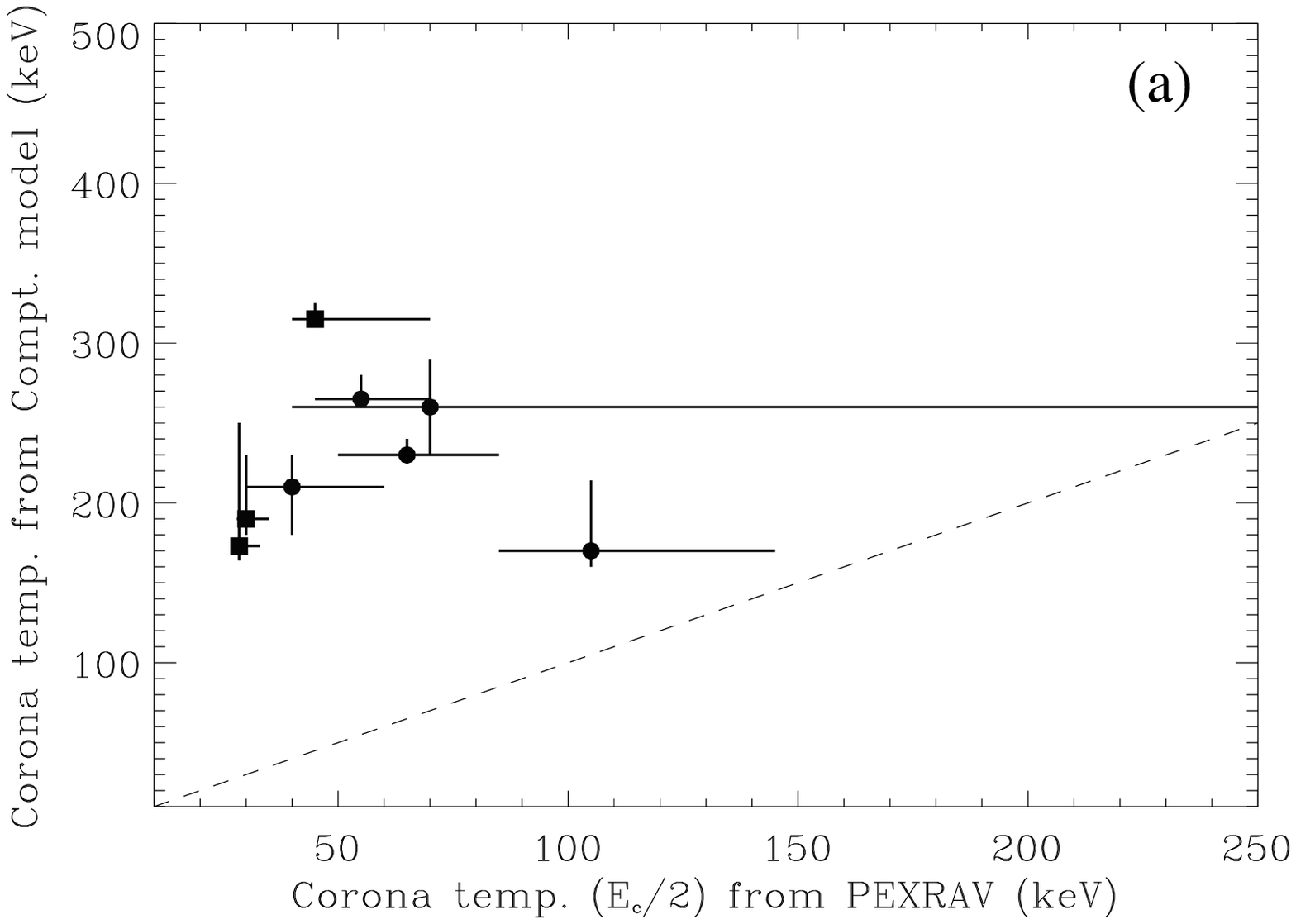}{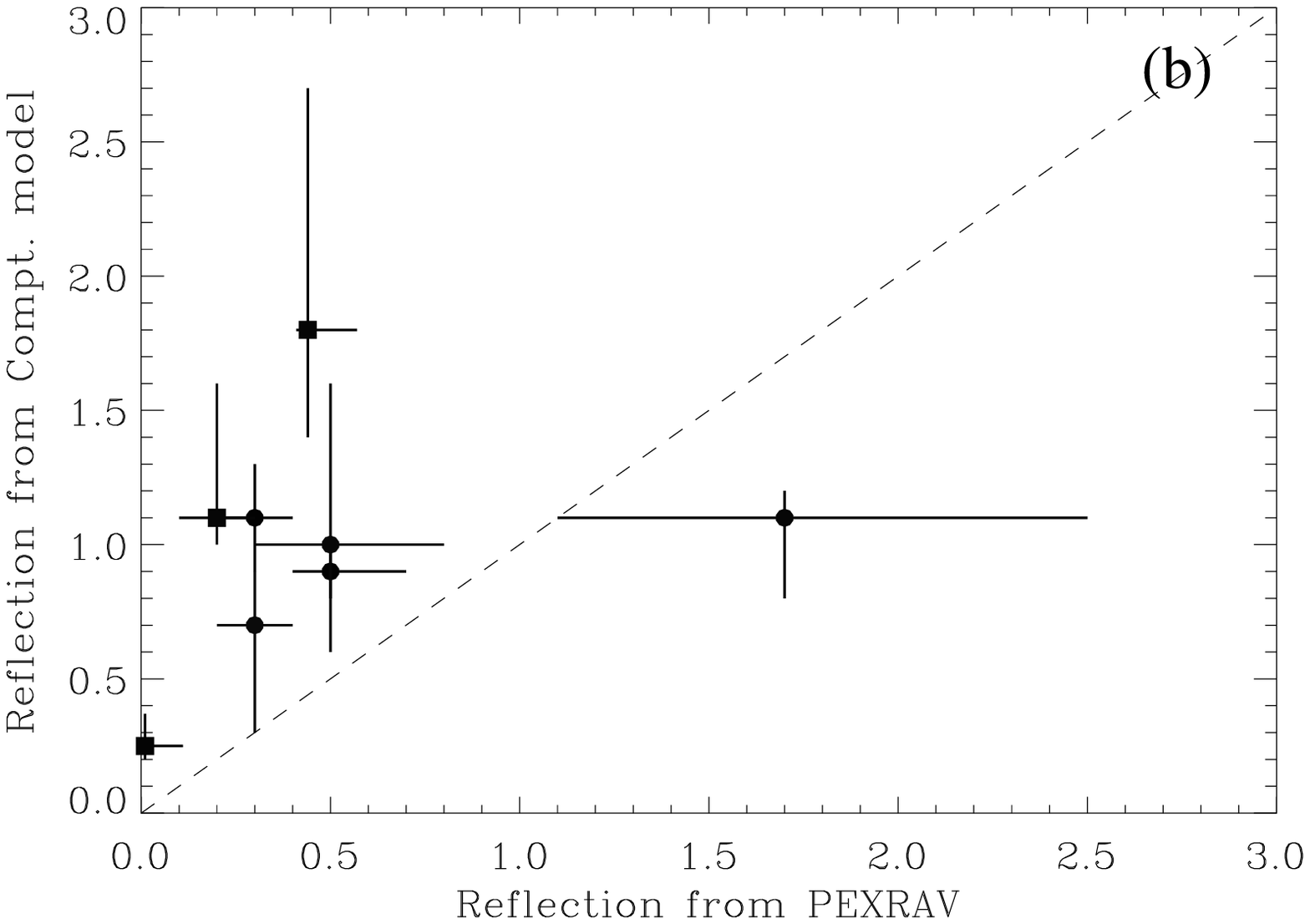}
\caption{}
\end{figure}

\clearpage

\pagestyle{empty}
\begin{figure}[h]
\plottwo{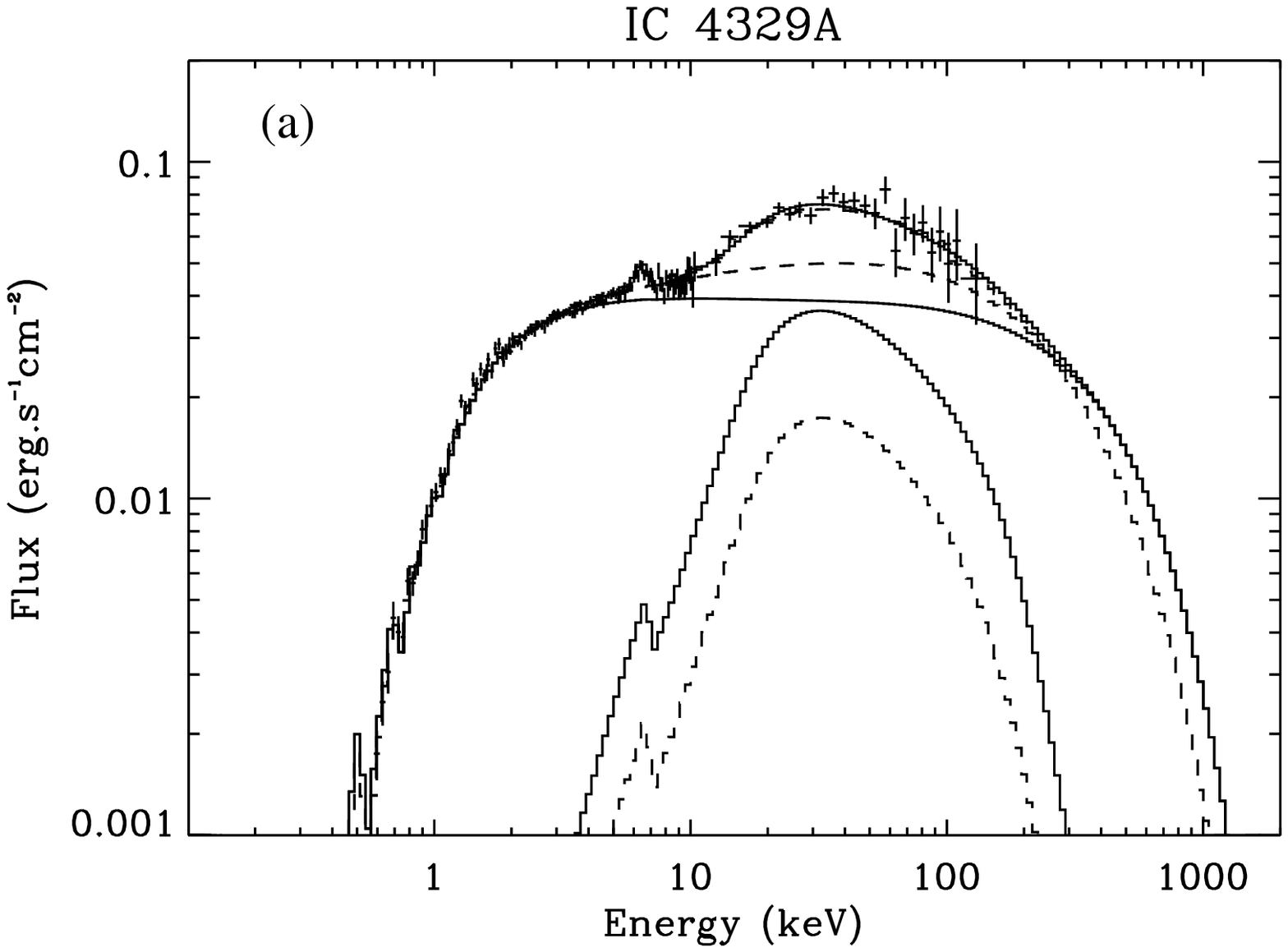}{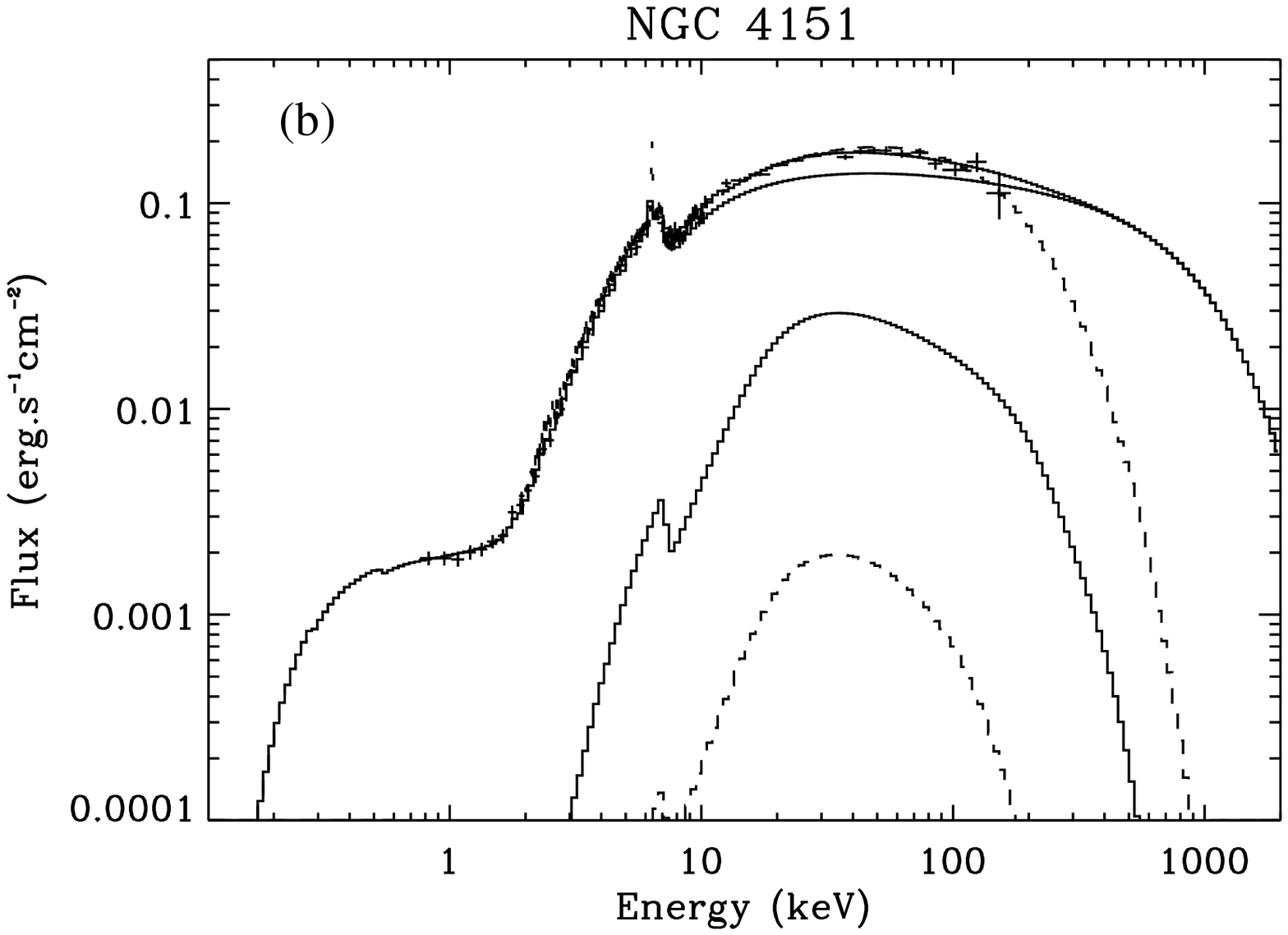}
\caption{}
\end{figure}

\clearpage

\begin{figure}[h]
\plottwo{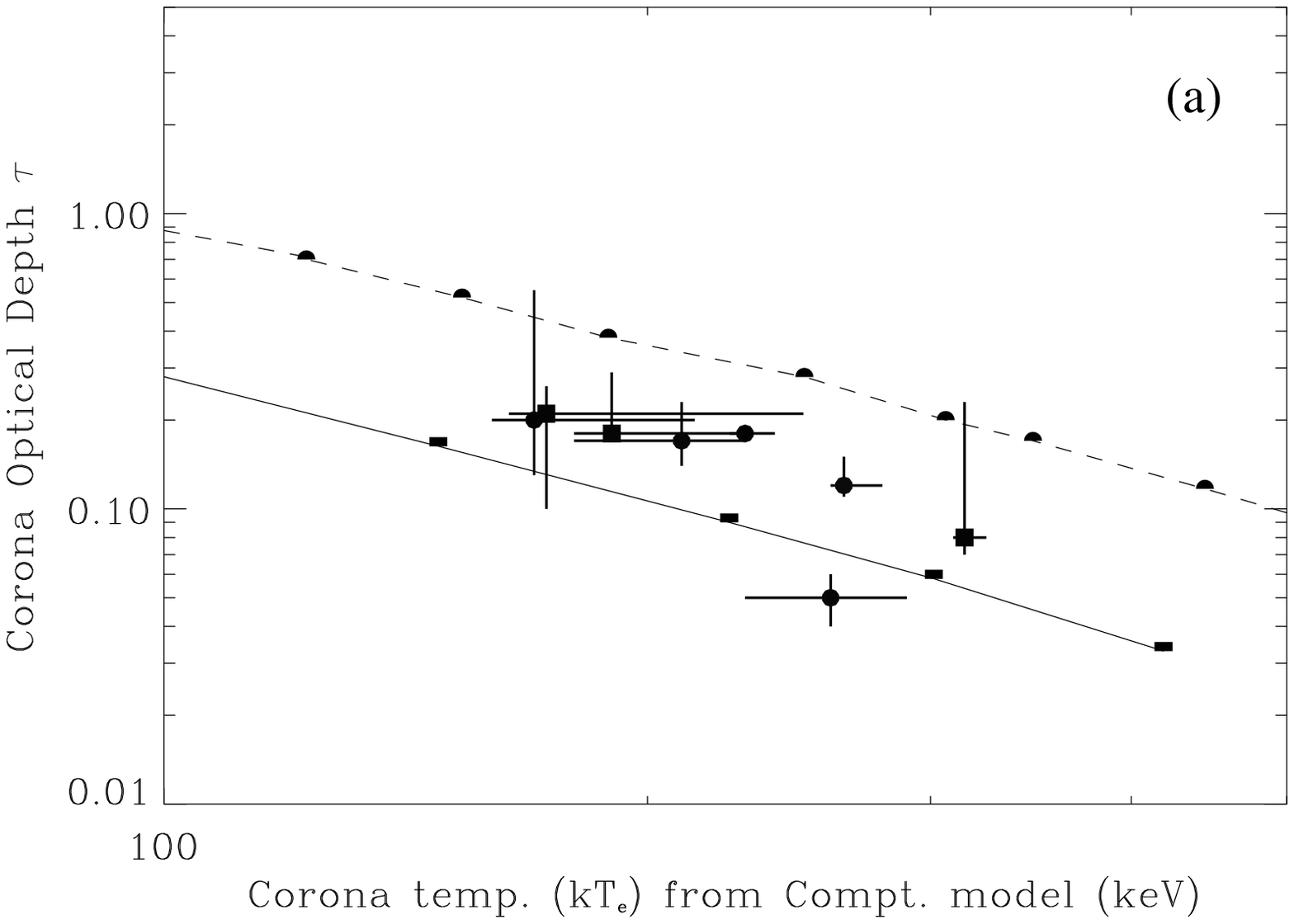}{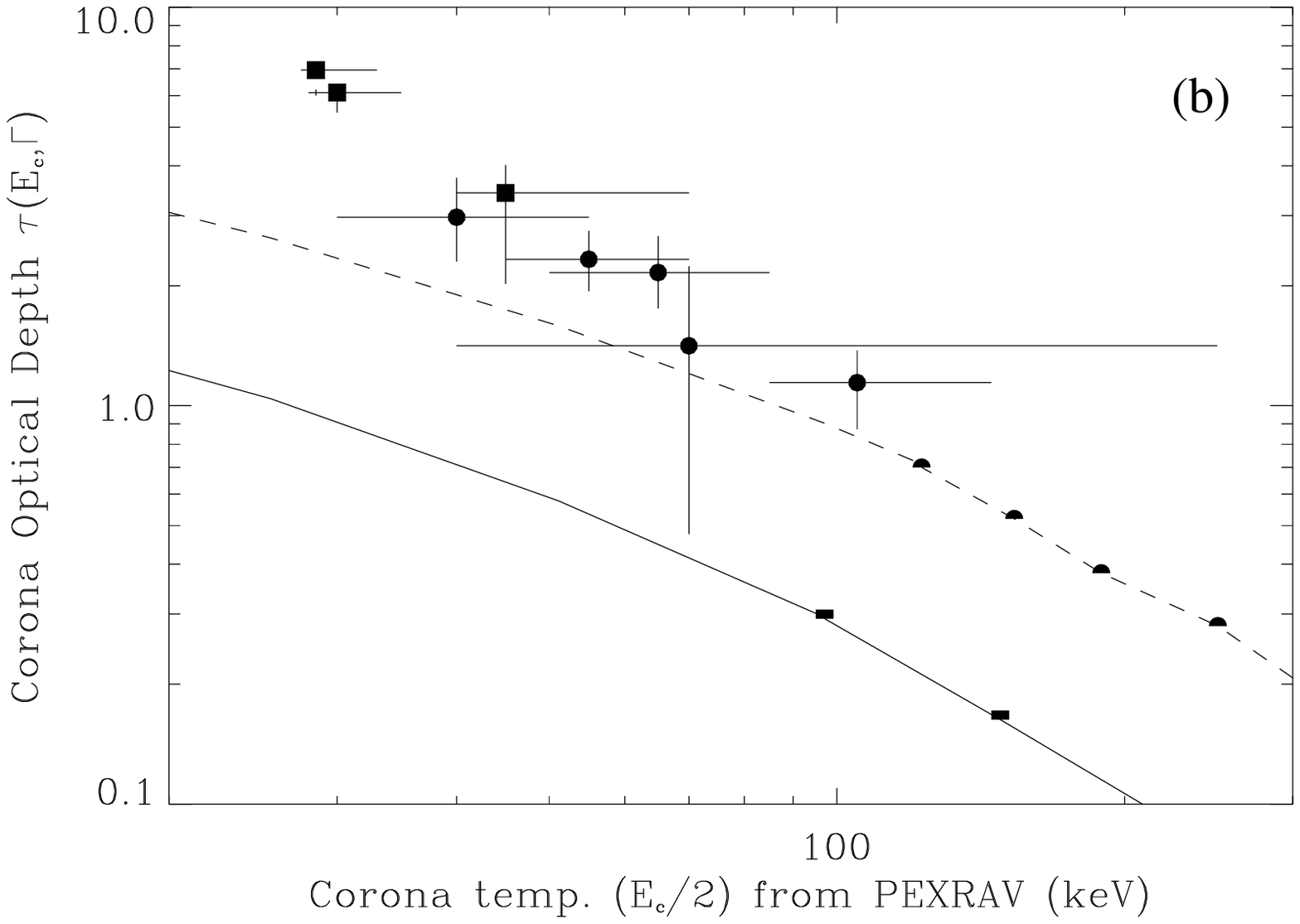}
\caption{}
\end{figure}

\clearpage

\begin{figure}[h]
\plotone{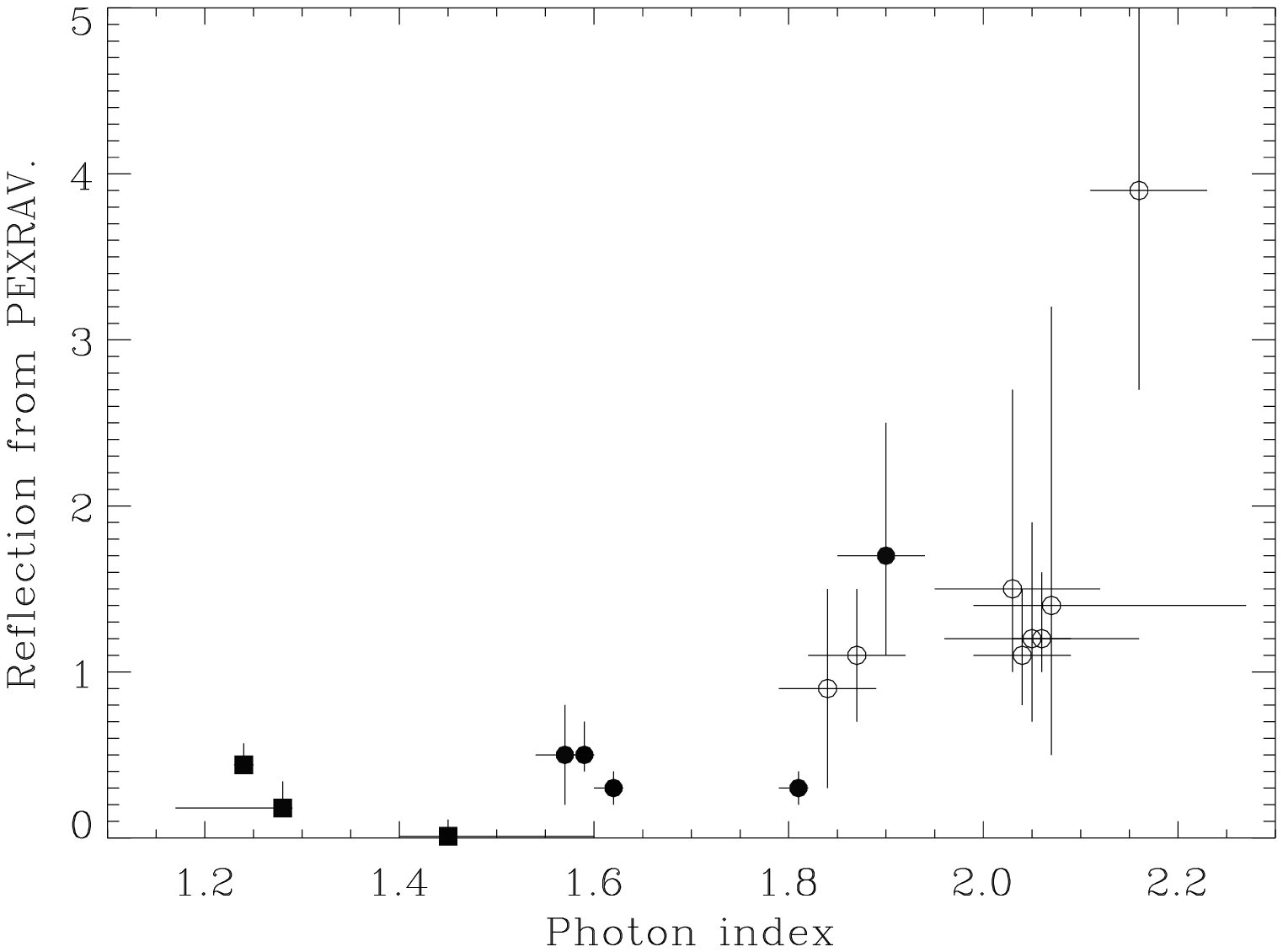}
\caption{}
\end{figure}

\clearpage

\begin{figure}[h]
\plotone{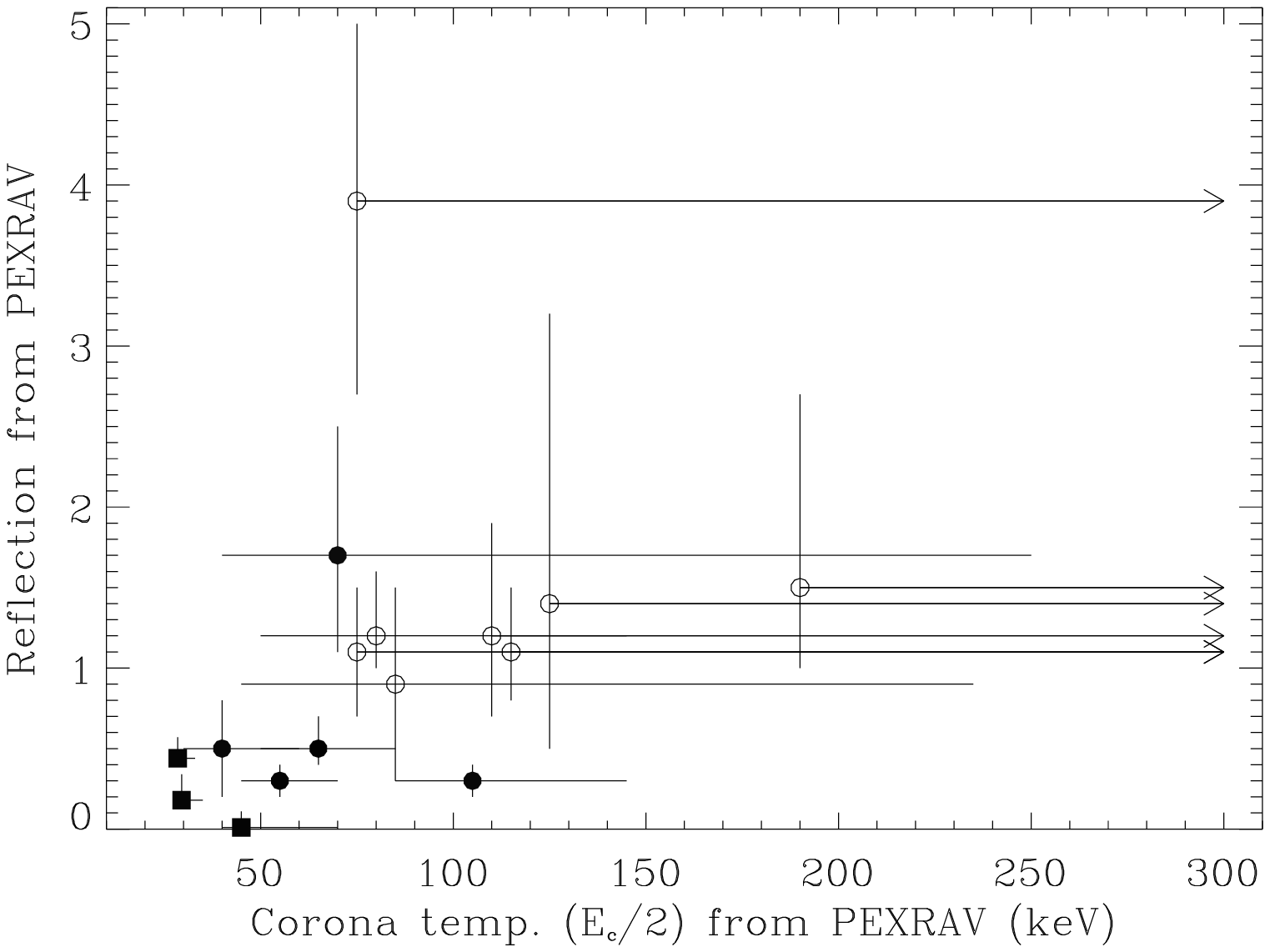}
\caption{}
\end{figure}

\clearpage

\begin{figure}[h]
\plottwo{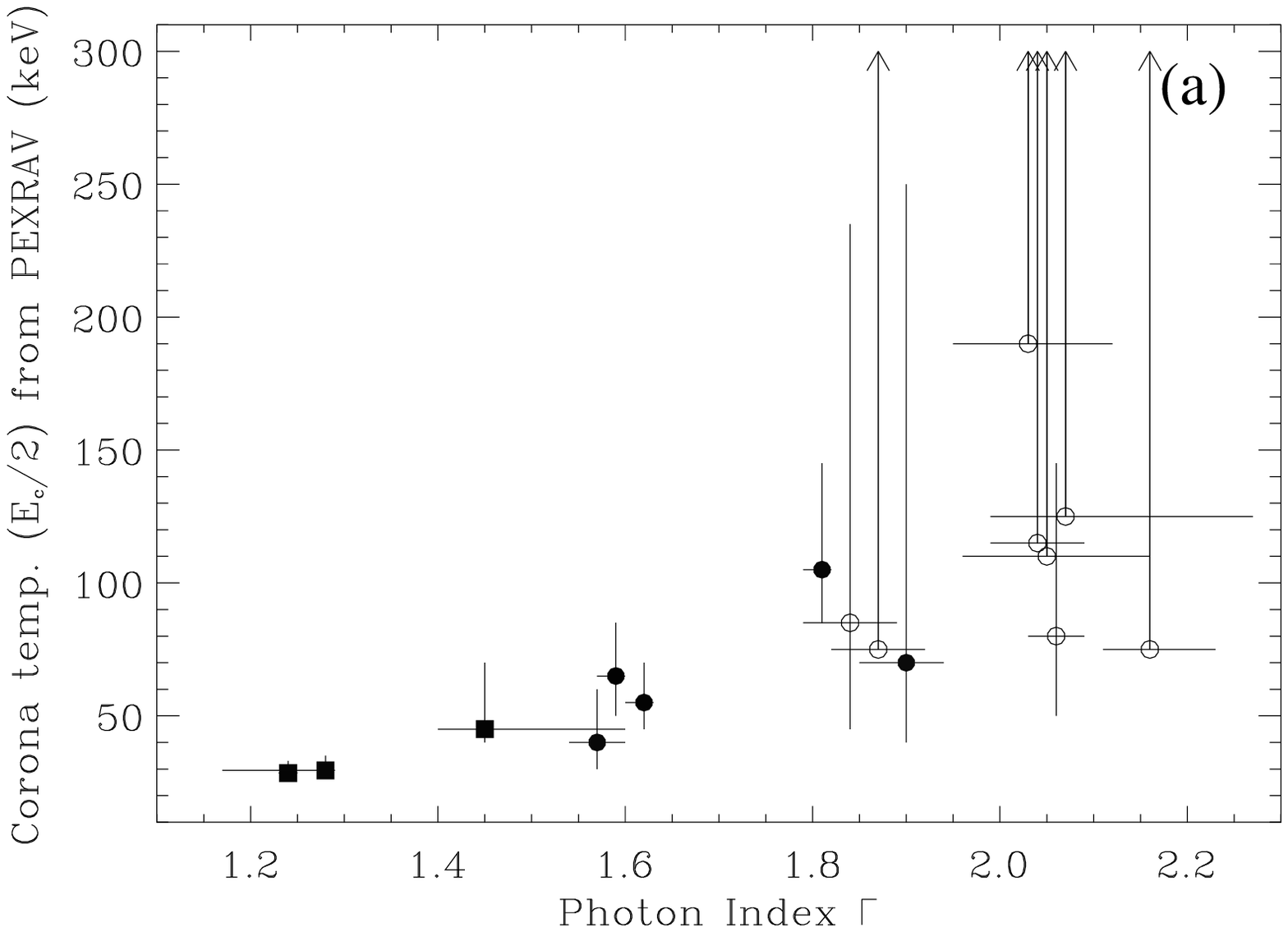}{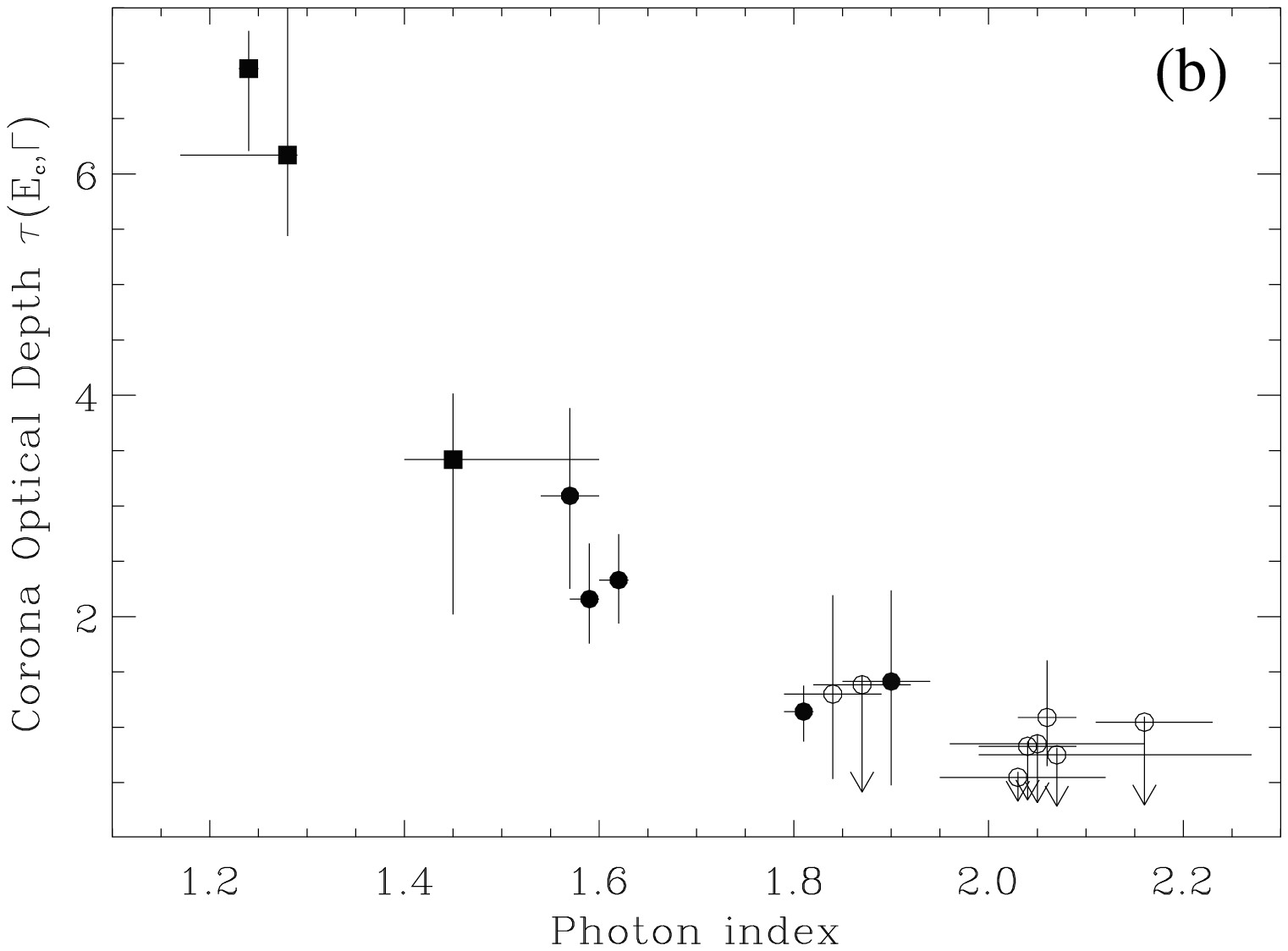}
\caption{}
\end{figure}

\clearpage

\begin{figure}[h]
\plotone{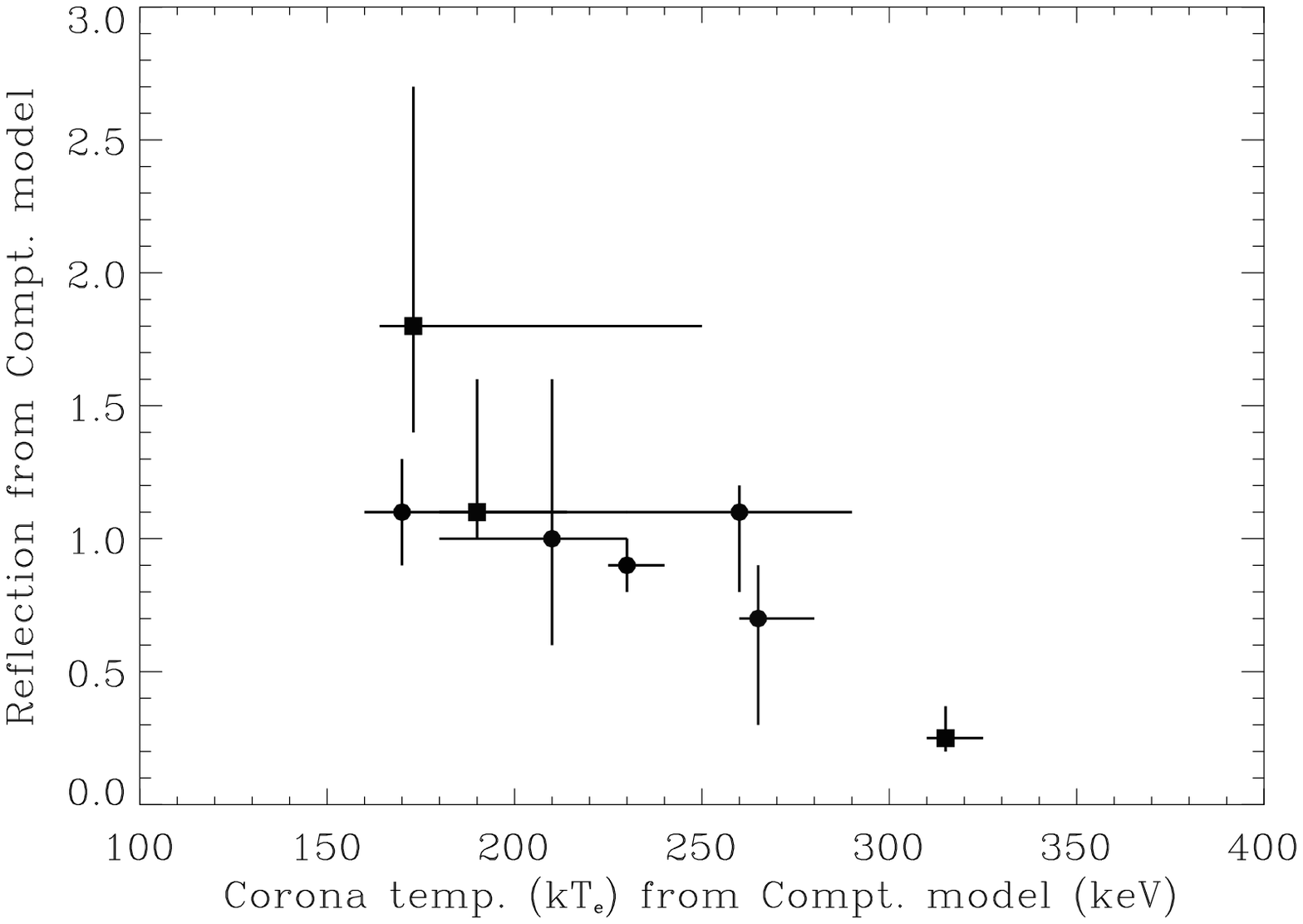}
\caption{}
\end{figure}

\clearpage

\begin{figure}[h]
\plotone{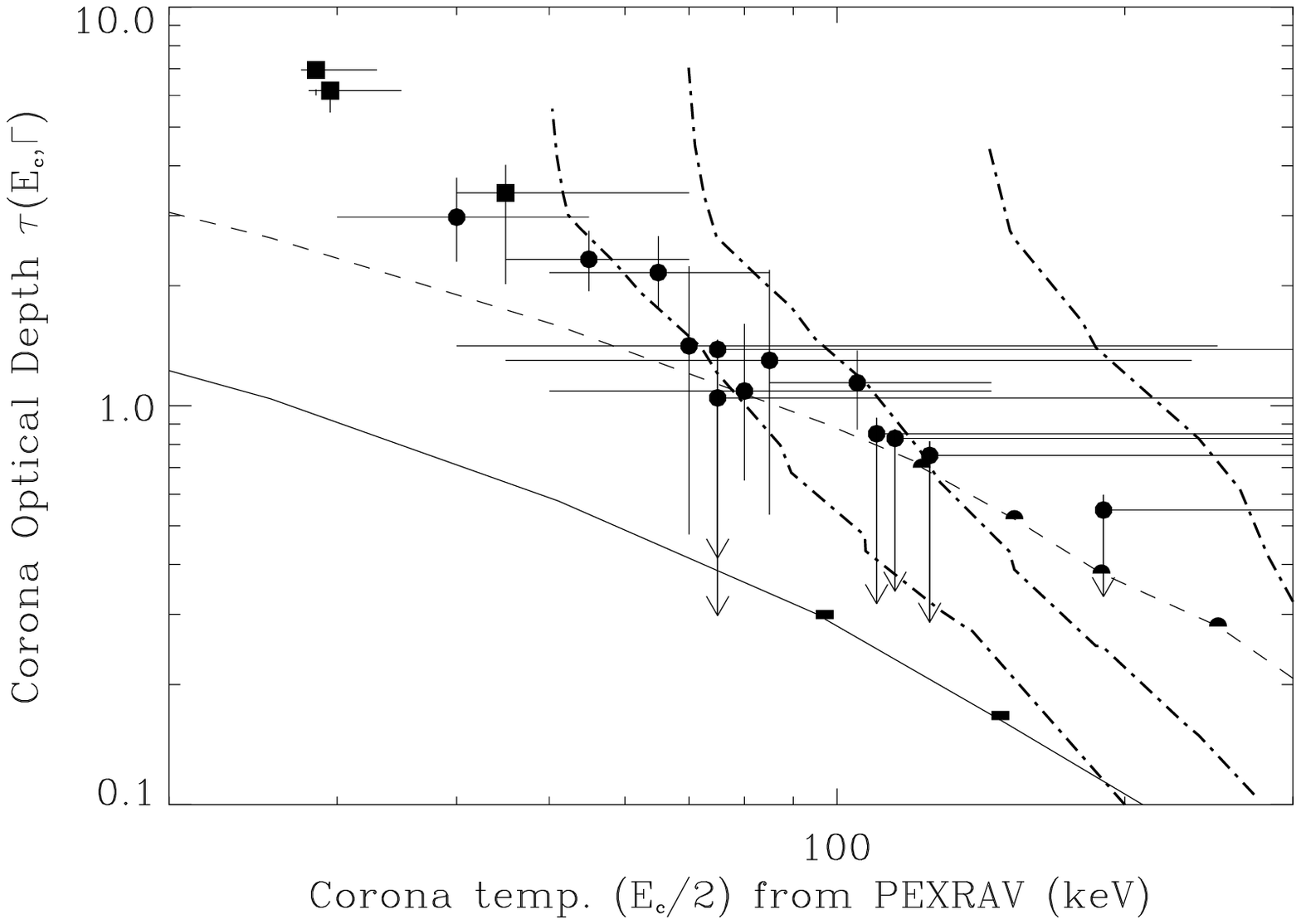}
\caption{}
\end{figure}

\end{document}